# Emergent helical texture of electric dipoles

Dmitry D. Khalyavin*[1], Roger D. Johnson[2,3], Fabio Orlandi[1], Paolo G. Radaelli[3], Pascal Manuel[1], Alexei A. Belik[4]

[1]ISIS Facility, Rutherford Appleton Laboratory, Harwell Oxford, Didcot OX11 0QX, United Kingdom

[2]Department of Physics and Astronomy, University College London, London, WC1E 6BT, United Kingdom

[3]Department of Physics, University of Oxford, Clarendon Laboratory, Parks Road, Oxford, OX1 3PU, United Kingdom

[4]International Center for Materials Nanoarchitectonics (WPI-MANA), National Institute for Materials Science (NIMS), Namiki 1-1, Tsukuba, Ibaraki 305−0044, Japan

*Correspondence to: dmitry.khalyavin@stfc.ac.uk

**Long-range ordering of magnetic dipoles in bulk materials gives rise to a broad range of magnetic structures, from simple collinear ferromagnets and antiferromagnets, to complex magnetic helicoidal textures stabilized by competing exchange interactions. In contrast, in the context of dipolar order in dielectric crystals, only parallel (ferroelectric) and antiparallel (antiferroelectric) collinear alignments of electric dipoles are typically considered. Here, we report an observation of incommensurate helical ordering of electric dipoles by light hole-doping of the quadruple perovskite $BiMn_7O_{12}$. In analogy with magnetism, the electric dipole helicoidal texture is also stabilized by competing instabilities. Specifically, orbital ordering and lone electron pair stereochemical activity compete, giving rise to phase transitions from a non-chiral cubic structure, to an incommensurate electric dipole and orbital helix, via an intermediate density wave.**

The long-range ordering of magnetic and/or electric dipoles are canonical causes of phase transitions in crystalline materials, and they are both associated with a variety of functional properties. In spite of the fundamentally different nature of the interactions between magnetic and electric dipoles, there are many similarities between the two types of phase transition, as first mentioned by Joseph Valasek *(1)* in 1922. In particular, many materials display parallel dipole alignments, like in ferromagnets and ferroelectrics, or antiparallel dipole alignments, like in antiferromagnets and antiferroelectrics. First predicted theoretically by Louis Néel in 1948 *(2)*, antiferromagnetic spin ordering was experimentally confirmed in MnO one year later using the newly established neutron diffraction technique *(3)* – a discovery that led to rapid experimental and theoretical developments *(4)*. Antiferroelectricity was first proposed in PbZrO$_3$ by Sawaguchi, Maniwa and Hoshino in 1951 *(5)*, based on the antiparallel displacements of Pb established by X-ray diffraction. Although by far the most common, collinear ferro- and antiferro-dipolar orderings (as well as their simple combinations like ferri- and canted structures) are not the only possibilities. In the late 1950s and early 1960s, Koehler et al *(6, 7)* and Yoshimori *(8)* discovered the first examples of non-collinear helical magnetic ordering in Ho and other rare earth metals, as well as in rutile MnO$_2$. Nowadays, many exotic orderings of magnetic dipoles are known, where spins gradually rotate creating complex patterns whose period is often incommensurate with the crystal structure. Such spin textures, characterized by additional macroscopic quantities such as chirality and polarity, play a central role in the physics of multiferroic materials, frustrated magnetism, skyrmions and magnetic domain walls. By analogy, one could expect that helical ordering in materials containing electric dipoles should also occur. However, such electric dipole textures have been challenging to observe experimentally in a bulk material. In the present work, we report an example of incommensurate helical ordering of electric dipoles in the lightly doped quadruple perovskite $BiCu_xMn_{7-x}O_{12}$. In addition to completing the analogy between ordering of magnetic and electric dipoles, our discovery demonstrates that many principles of non-collinear magnetism can be adopted to understand the properties of dielectric materials.

The parent compound $BiMn_7O_{12}$ is a so-called quadruple perovskite with $Mn^{3+}$ on both A and B perovskite sites, $[BiMn^{3+}_3]_A[Mn^{3+}_4]_BO_{12}$. It combines two well-known electronic instabilities, leading to a complex structural behavior *(9)*: the stereo-chemical activity of $Bi^{3+}$ caused by the $6s^2$ lone-pair electrons and the Jahn-Teller (JT) instability caused by the degenerate $e_g$ electronic states of octahedrally coordinated $Mn^{3+}$. The former promotes off-centric cation displacements and the formation of local electric dipoles *(10)*, and is hence responsible for the ferroelectric properties of many Bi-containing perovskite oxides *(11-13)*. The latter often results in a structural phase transition characterized by a coherent, spontaneous distortion of polyhedra coordinating $Mn^{3+}$, as for instance in the case of $LaMnO_3$ *(14)*. At high temperatures, $BiMn_7O_{12}$ adopts a cubic non-ferroelectric structure, with space group (SG) *Im-3*, whereas on cooling it undergoes three distinct structural transitions at $T_{JT} \sim 608$ K, $T_C \sim 460$ K and $T_S \sim 290$ K. The highest-temperature transition at $T_{JT}$ (to SG *I2/m*) is driven by cooperative JT distortions, leading to a B-site orbital pattern identical to the prototype manganite $LaMnO_3$ *(14)*. At the Curie temperature $T_C$, $BiMn_7O_{12}$ becomes ferroelectric owing to ferro-dipolar ordering of the Bi lone pairs, whereas $T_S$ is a crystallographic transition between the higher-temperature (SG *Im*) and ground state (SG *P*1) ferroelectric phases. This transition can be interpreted as an electric dipole re-orientation at which the electric polarization (initially confined to be within the *ac* plane) rotates to a general direction *(15)*. Throughout both ferroelectric phases the $LaMnO_3$-type orbital order remains largely unchanged.

As is the case for other manganites *(16, 17)*, a small amount of hole doping achieved, for example, by the replacement of A site $Mn^{3+}$ with $Cu^{2+}$, is expected to tune both instabilities because it affects both the JT ordering (doped holes are localized on B sites *(18, 19)* such that $Mn^{3+}$ tends towards non JT active $Mn^{4+}$) and the immediate coordination of $Bi^{3+}$. Similar to the undoped compound, $BiCu_{0.1}Mn_{6.9}O_{12}$ is cubic at high temperatures (structural parameters for this phase as well as for the phases discussed below are summarized in tables S1 to S5 and figures S1 to S8 of *(20)*), and undergoes a JT order phase transition to the non-polar monoclinic structure with *I2/m* SG at a slightly lower $T_{JT}$ ~ 560 K (Fig. 1A and fig. S3) accompanied by the same $LaMnO_3$-type orbital order.

Compared to $BiMn_7O_{12}$, the nature of the two lower-temperature phase transitions in $BiCu_{0.1}Mn_{6.9}O_{12}$ is notably different (Fig. 1A and figs. S4 to S6): the Curie and structural transitions are replaced by transitions to two incommensurately modulated phases at $T_{I1}$ ~ 435 K and $T_{I2}$ ~ 390 K, both of which will be described below. The transitions are strongly first order and are associated with large thermal hysteresis (fig. S9). In both modulated phases the incommensurate satellite reflections are very strong. These sudden structural changes at such low doping strongly suggest that the modulation can be associated with reconstruction of the primary electronic instabilities, namely, ferroelectric dipole ordering and/or orbital polarization.

Crystallographic analysis performed at $T$ = 427 K (below $T_{I1}$) revealed that the structure is metrically triclinic and that the modulation vector is incommensurate $k_{HT}$ = (-0.0037(8), 0.026(1), 0.1233(8)) – approximately parallel to the body diagonal of the pseudo-cubic cell. Structural refinements performed in the *R-1(αβγ)0* superspace group (SSG) established that the Bi cations predominantly shift along the a-axis (former cubic [0-11]$_C$ direction) (Fig. 2A) and adopt the largest modulation amplitude of all atomic species. This confirms that the modulation is related to the ordering of lone electron pairs, which in the parent material, $BiMn_7O_{12}$, induces ferroelectricity by adopting parallel ordering. In contrast, in the triclinic phase of $BiCu_{0.1}Mn_{6.9}O_{12}$ the Bi displacements are aligned on a fixed axis, but their amplitude is modulated giving zero net ferroelectric polarization. This electric dipole ordering strongly resembles the incommensurate structure of magnetic moments in a spin density wave (SDW) – a feature of many magnetic systems with frustrated exchange interactions. Our triclinic 'Dipole Density Wave' (DDW) structure is stable only in a narrow temperature range below the phase transition – another element of similarity with SDWs, which are almost always entropically stabilized.

The low-temperature phase of $BiCu_{0.1}Mn_{6.9}O_{12}$, below $T_{I2} \sim 390$ K and down to the ground state, is metrically trigonal and heavily modulated along the c-axis; $\boldsymbol{k_{LT}} = (0,0,\gamma)$ with $\gamma = 0.1046(5)$ at $T = 380$ K. This modulation vector is on the $\Lambda$ line of symmetry and is oriented exactly along the body diagonal of the pseudo-cubic cell ($\boldsymbol{k_{LT}} = (2/3\gamma, 2/3\gamma, 2/3\gamma)$ in the $Im$-3 setting). A systematic test of all isotropy subgroups of $Im$-3 associated with this type of modulation *(21-23)* revealed that only the non-centrosymmetric SSG $R3(00\gamma)t$ can provide an excellent refinement quality for both neutron (Fig. 1B) and X-ray (fig. S5) diffraction patterns. The obtained crystal structure comprises a distinct pattern of modulated distortions (Fig. 1C). The largest refined atomic displacements are, once again, associated with Bi and lie in the $ab$ plane of the trigonal cell, and all have the same amplitude of 0.41(1) Å at $T = 300$ K. This value is extremely large: it exceeds the ferroelectric displacements of the cations in the undoped $BiMn_7O_{12}$ (~ 0.37 Å) *(9, 24)* and $BiFeO_3$ (~ 0.40 Å) *(25)*.

In the lower temperature phase of $BiCu_{0.1}Mn_{6.9}O_{12}$ ($T < 390$ K), the displacement vectors rotate perpendicular to the modulation wavevector propagating along the c-axis (Fig. 1C and Fig. 2, B and C), and are therefore the structural analogue of a magnetic incommensurate proper helix.

Thus, in this phase, the lone electron pairs of $Bi^{3+}$ cations are ordered into helical structure, making $BiCu_{0.1}Mn_{6.9}O_{12}$ a "textbook" example of a helical texture of electric dipoles. In general, the order parameter of a structural or magnetic helical phase can be defined as a mixed product $\sigma_s = \mathbf{k}_{LT} \cdot [\mathbf{r}_i \times \mathbf{r}_j]$ or $\sigma_m = \mathbf{k}_m \cdot [\mathbf{S}_i \times \mathbf{S}_j]$, where $\mathbf{r}_i$ and $\mathbf{r}_j$ ($\mathbf{S}_i$ and $\mathbf{S}_j$) are atomic displacements (spins) in adjacent unit cells along the propagation vector $\mathbf{k}_{LT}$ ($\mathbf{k}_m$) (Fig. 2, B and C). In both cases, this quantity is a time-reversal-even pseudo-scalar, which in the present case measures the modulus and sign of the structural chirality (or helicity). Once again, the parallel with the magnetic counterparts is particularly apparent in the sequence of phase transitions we have observed in $BiCu_{0.1}Mn_{6.9}O_{12}$ (paraelectric – DDW – helical), which is completely analogous to that of frustrated magnets such as the prototypical Type-II multiferroic $TbMnO_3$ (paramagnetic – SDW – cycloidal) *(26, 27)*.

One striking implication is that the $BiCu_xMn_{7-x}O_{12}$ series provides a distinctive example of emergent structural chirality acting as a macroscopic order parameter (ferrochirality). In most Type-II multiferroics including the prototypes $TbMnO_3$ *(26-28)* and $CaMn_7O_{12}$ *(29, 30)*, spin (vector) chirality appears below a magnetic phase transition and is coupled to improper ferroelectricity through a suitable free-energy invariant. In $TbMnO_3$, this term describes a direct coupling between vector spin chirality and the electric polarization, and is allowed only in the weakly ferroelectric cycloidal phase, and not in the high-temperature, non-polar SDW phase *(27, 28)*. Unlike the case of $TbMnO_3$, where the magnetic chirality and electric polarization are directly coupled, multiferroicity in $CaMn_7O_{12}$ requires the presence of an additional axial structural distortion $A$ (known as ferro-axiality) *(29-31)* to form the free-energy invariant $\sigma_m A P$. The above analysis leads us to the prediction that improper (weak) ferroelectricity should also be induced in $BiCu_{0.1}Mn_{6.9}O_{12}$ by precisely the same mechanism, as the free-energy invariant $\sigma_s A P$ (with the spin chirality $\sigma_m$ being replaced by the structural chirality $\sigma_s$) is allowed in the low-temperature $R3(00\gamma)t$ phase of BiCu$_{0.1}$Mn$_{6.9}$O$_{12}$. We have tested this prediction by performing ferroelectric (*P-E*) hysteresis loops measurements (fig. S10) at liquid nitrogen temperature (77 K). It is well-known that standard (*P-E*) measurements can be compromised by factors unrelated to the intrinsic polarization switching, such as leakage currents *(32)*. We have therefore employed the so called Positive-Up-Negative-Down (PUND) method, which has been developed to distinguish true, intrinsic ferroelectric switching from artificial responses due to current leakage *(33)*. The resultant (*P-E*) loop is consistent with the phenomenological prediction of weak improper ferroelectric polarization in the helical phase, with |***P***| being 4000 times smaller than the value of polarization estimated for the parent compound BiMn$_7$O$_{12}$ *(24)*. By symmetry, ***P*** should be directed along the c-axis, although this could not be established for our polycrystalline samples. We note that the coexistence of ferroelectric and chiral orders is also known in smectic phases of liquid crystals with chiral molecules *(34)*. These substances however are non-crystalline with only directional order of anisotropically shaped molecules, and hence they relate to substantially different underlying physics and materials properties.

In multiferroics it has been shown that magnetic domains with different spin chirality can be controlled by external electric fields *(35)*. In complete analogy with this, we predict that the structural chirality domains in BiCu$_{0.1}$Mn$_{6.9}$O$_{12}$ should also be controllable by an electric field, given that the polarization takes opposite directions in these domains (Fig. 2, B and C). Should this be confirmed by single crystal or epitaxial thin films measurements, BiCu$_{0.1}$Mn$_{6.9}$O$_{12}$ would be a unique system in which structural chirality could be permanently switched by a static electric field. This effect would be essentially different to the electro-gyration effect, in which the induced chirality has no associated hysteresis. Importantly, this functionality could be amenable to a variety of perspective applications based on materials properties associated with structural chirality *(36, 37)*, such as optical activity (a rotation of the polarization of linearly polarized light) and circular dichroism (differential absorption of left- and right-handed polarized light), particularly because the chiral phase in $BiCu_xMn_{7-x}O_{12}$ is stable up to temperatures well above room temperature.

One final question concerns the nature of orbital ordering in $BiCu_{0.1}Mn_{6.9}O_{12}$, and its relation with structural chirality. This can be established by inspecting the oxygen octahedra coordinating the B-site Mn ions in the different phases, given that differential occupancy of the $e_g$ electronic orbitals produces unequal B-O distances within the $BO_6$ octahedra. This can be quantified using the mixing angle formalism proposed by Goodenough *(38)*. The mixing angle, $\theta$, defines an admixture between $3z^2-r^2$ and $x^2-y^2$ orbital states within an 'orbital plane', through the relation $\tan\theta = \sqrt{3}(d_{B-Oy} - d_{B-Oz})/(2d_{B-Ox} - d_{B-Oy} - d_{B-Oz})$, where $d_{B-O}$ are B-O bond lengths along $x$, $y$ or $z$ directions of the local Cartesian coordinate system as shown in Fig. 3, A to C and E note that in our analysis we employ oxygen-oxygen distances, $d_{O-O} = 2 <d_{B-O}>$, to reduce the experimental uncertainty). As previously stated, the monoclinic phase just below $T_{JT}$ adopts the $LaMnO_3$ - type planar orbital ordering characterized by a commensurate pattern of $BO_6$ octahedral distortions, which gives close to fully polarized orbital states for all manganese ions (Fig. 3A). By comparison, in the triclinic DDW and the trigonal helical phases, the incommensurate atomic displacements modulate the $BO_6$ octahedral distortions giving rise to an incommensurate modulation of the orbital occupancy (i.e. an Orbital Density Wave, ODW). The closely related quadruple perovskite $CaMn_7O_{12}$, also displays an ODW, with the orbital occupancy gradually changing between $3x^2-r^2$ and $3y^2-r^2$ along the modulation direction *(30)*. In the triclinic modulated structure of $BiCu_{0.1}Mn_{6.9}O_{12}$, three out of the four symmetry distinct Mn B-sites (Mn4, Mn5, and Mn6) display a $CaMn_7O_{12}$ - type ODW (albeit in different directions, Fig. 3B and fig. S4), whereas the Mn7 site maintains the fully polarized, unmodulated orbital state inherited from the higher-temperature monoclinic phase. The ODW in the trigonal helical phase involves two symmetry independent Mn B-sites, Mn2 and Mn3 (Fig. 3C and fig. S6). The latter is located on a 3-fold axis of symmetry and the mixing angle $\theta$ for this site takes all values between 0 and 360 degrees through the period of the modulation (Fig. 3, D and E). This implies that the orbital state of the Mn3 continuously rotates in the orbital plane either clockwise or counter clockwise, following the rotation of the Bi-displacements, and giving rise to a chiral ODW (Fig. 3, E and F). The orbital occupancies on the Mn2 sites, which can be considered as triangular motifs inter-related by 3-fold symmetry, are also modulated and alternate between $3z^2-r^2$ and $z^2-x^2$ states (and the equivalent states obtained by 3-fold rotation). The phase of the modulation is not uniform on the triangle of sites, as it is in $CaMn_7O_{12}$, but instead differs by $+/-2\pi/3$, also making the global Mn2 ODW chiral.

Thus, the orbital ordering in the hole-doped manganite $BiCu_{0.1}Mn_{6.9}O_{12}$ gradually evolves from the monoclinic $LaMnO_3$ - type, commensurate phase with nearly fully polarized orbital states but disordered Bi displacements, to the incommensurate trigonal phase with chiral ODW and helical (chiral) Bi dipole displacements. The evolution takes place via an intermediate incommensurate triclinic phase combining the non-chiral Bi DDW and the non-chiral Mn ODW. This remarkable structural behavior can be interpreted as a competition between the two primary electronic instabilities, one related to the lone electron pair of $Bi^{3+}$, and the other to the degeneracy of $e_g$ electronic states of the octahedrally coordinated B-site $Mn^{3+}$ - both requiring different types of structural distortions. This again underscores the analogy with magnetism, where the non-collinear spin textures appear as a result of competing exchange interactions. The apparent similarity between non-collinear electric and magnetic dipole orderings demonstrates that the formation of chiral incommensurate structures is in fact a universal mechanism that allows a coexistence of competing degrees of freedom. An immediate implication of the presently observed chiral electric dipole and orbital textures lies in our understanding of ferroelectric domain walls, whereby a growing number of experiments have shown that the ordering phenomenon in ferroelectric domain walls is far more complicated than had been commonly believed before *(39-42)*. In this sense, the BiCu$_x$Mn$_{7-x}$O$_{12}$ helical dipole phase can be thought of as the extended version of a ferroelectric Bloch domain boundary. In addition, by analogy with magnetism, more complex but closely related chiral objects such as electric skyrmions are likely to be feasible as well. The magnetic skyrmion lattice is often described as a multi-*k* phase represented by the superposition of three helical spin textures and induced from a single-*k* state by external magnetic field *(43)*. One can expect that electric skyrmions can also be stabilized by external stimuli, or can be found in the domain boundaries of the BiCu$_x$Mn$_{7-x}$O$_{12}$ helical phase where three single-*k* states merge together. Crucially, the formation of chiral electric dipole textures has previously been discussed only in the context of complex artificial nanostructures *(44-46)* or domain walls *(39-42)*, with specific electrostatic boundary conditions or a delicate balance between electrostatic, elastic and lattice-mismatch strain energies. Our work demonstrates a distinct approach to stabilize these textures via competing electronic instabilities that is equally efficient for both bulk and low-dimensional structures. This opens an additional dimensionality in the design of materials, and suggests architectures for heterostructures, photonic crystals and superlattices that combine traditional ingredients with components that comprise layers of chiral electric and orbital textures.

**Acknowledgments:** We thank Dr. Yoshio Katsuya, Dr. Masahiko Tanaka, and Dr. Yoshitaka Matsushita of NIMS for their help with synchrotron X-ray and SEM data collections. The synchrotron radiation experiments were performed at SPring-8 with the approval of the NIMS Synchrotron X-ray Station (Proposal Numbers: 2016B4504 and 2017A4503). The authors also acknowledge the Science and Technology Facility Council for the provision of neutron beam time on the WISH instrument (RB1810030). **Funding:** R.D.J acknowledges support from a Royal Society University Research Fellowship. A.A.B. acknowledges partial support from grant JPJ004596 (ATLA, Japan). **Authors contributions:** D.D.K. and F.O. performed interpretation and refinement of the neutron and X-ray diffraction data collected by P.M. and A.A.B. at different facilities. R.D.J. and P.G.R. suggested the symmetry principles of coupling between structural chirality and macroscopic polarization and performed ($P$-$E$) measurements. A.A.B. developed the concept, carried out the materials synthesis, characterization and physical property measurements. All authors contributed to the preparation of the manuscript and to discussion of the results. **Competing interests:** The authors declare no competing interests. **Data and materials availability:** The experimental data used in the present study are available at Zenodo (47).

**Supplementary Materials:**

Materials and Methods

Figs. S1 to S10

Tables S1 to S5

References (*48-50*)

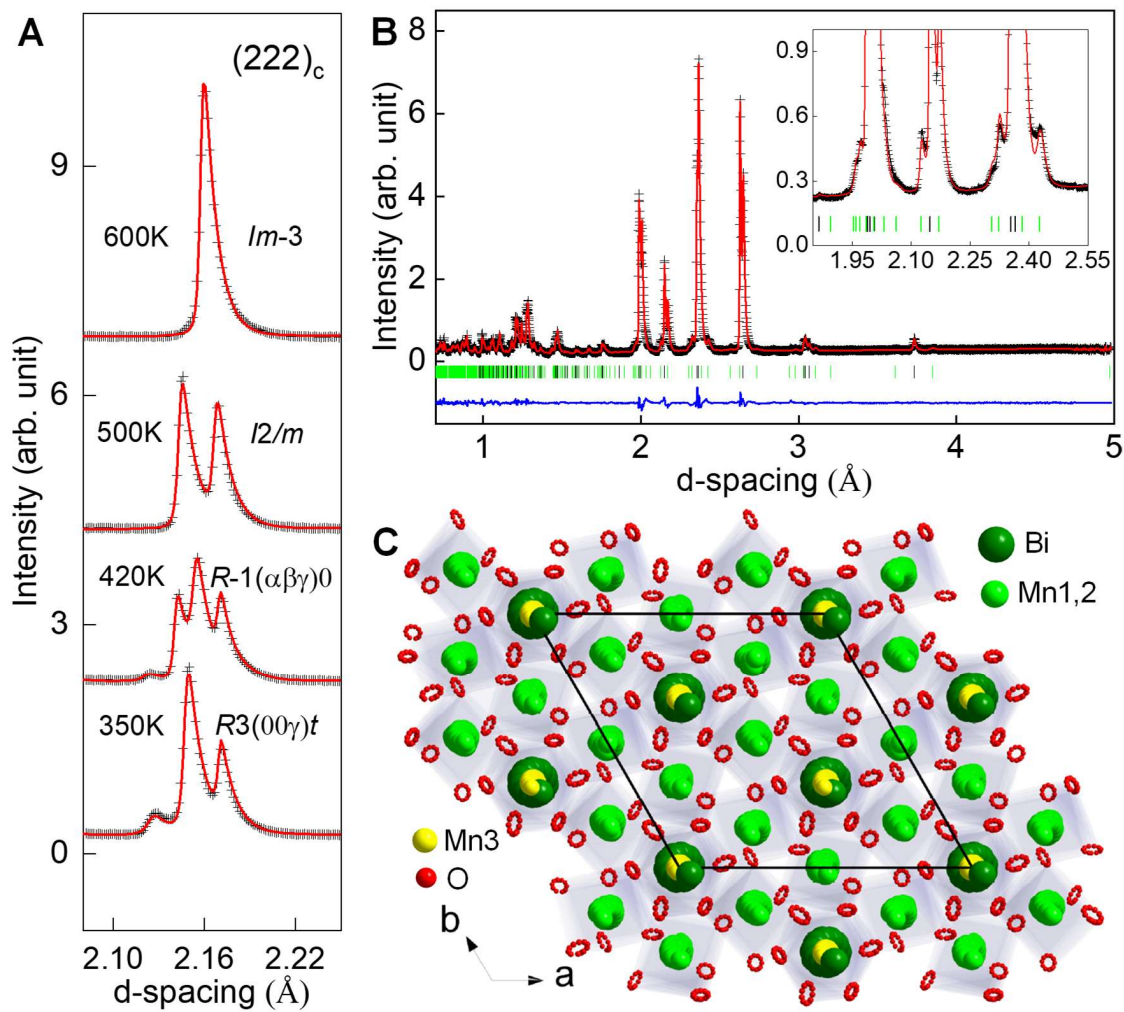

**Fig. 1. Crystal structure refinement.** (**A**) Thermal evolution of $(222)_C$ fundamental reflection measured by neutron diffraction in the different phases of $BiCu_{0.1}Mn_{6.9}O_{12}$. (**B**) Rietveld refinement of the neutron diffraction data collected at room temperature in the modulated helical phase. Cross symbols and solid red line represent experimental and calculated intensities, respectively, and the blue line below is the difference between them. The black tick marks indicate positions of the fundamental peaks, and the green tick marks indicate positions of the satellites. The corresponding structural parameters are listed in table S4. The inset shows a region where the strongest satellite reflections are observed in the modulated helical phase. (**C**) View along the c-

axis of the crystal structure of $BiCu_{0.1}Mn_{6.9}O_{12}$ in the modulated helical phase, demonstrating chiral atomic displacements.

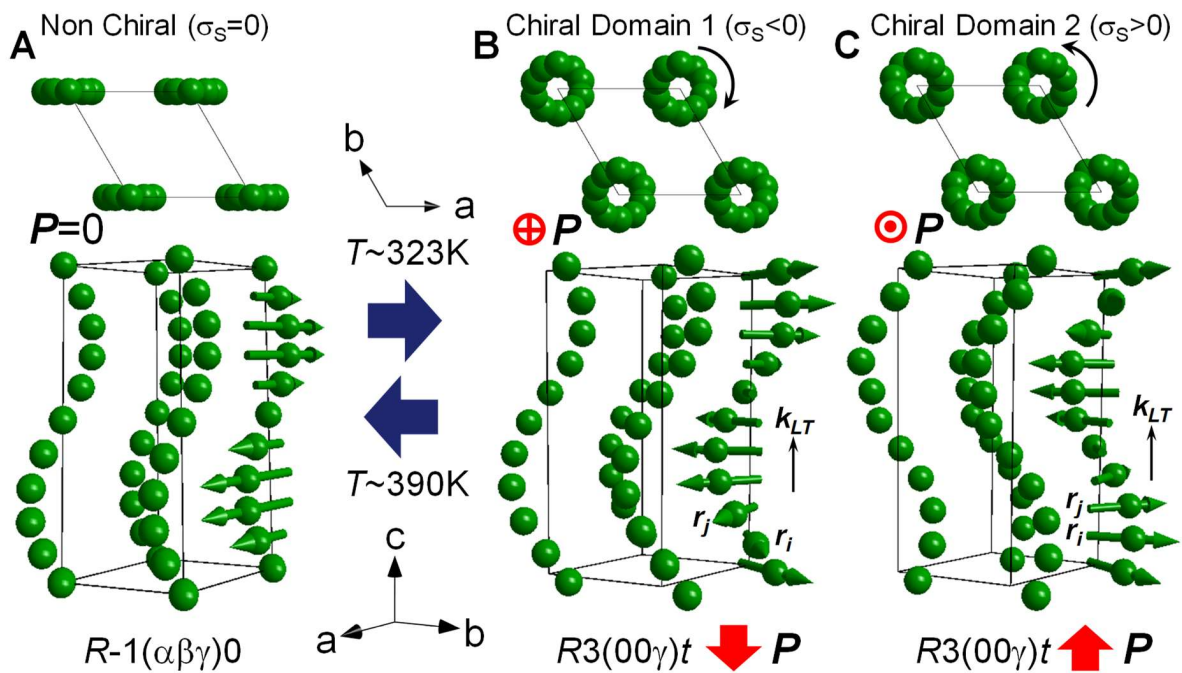

**Fig. 2. Modulated atomic displacements.** Schematic representation of the displacements of Bi in the **(A)** high temperature non-chiral and **(B, C)** low-temperature chiral modulated phases of $BiCu_{0.1}Mn_{6.9}O_{12}$. Atoms related by *R*-centering are omitted for clarity. The black lines indicate an approximately commensurate supercell. The displacement directions are shown by arrows whose length is proportional to the magnitude of the displacement. In the chiral phase, two domains with opposite chirality, and hence opposite direction of the weak macroscopic polarization ***P***, are shown.

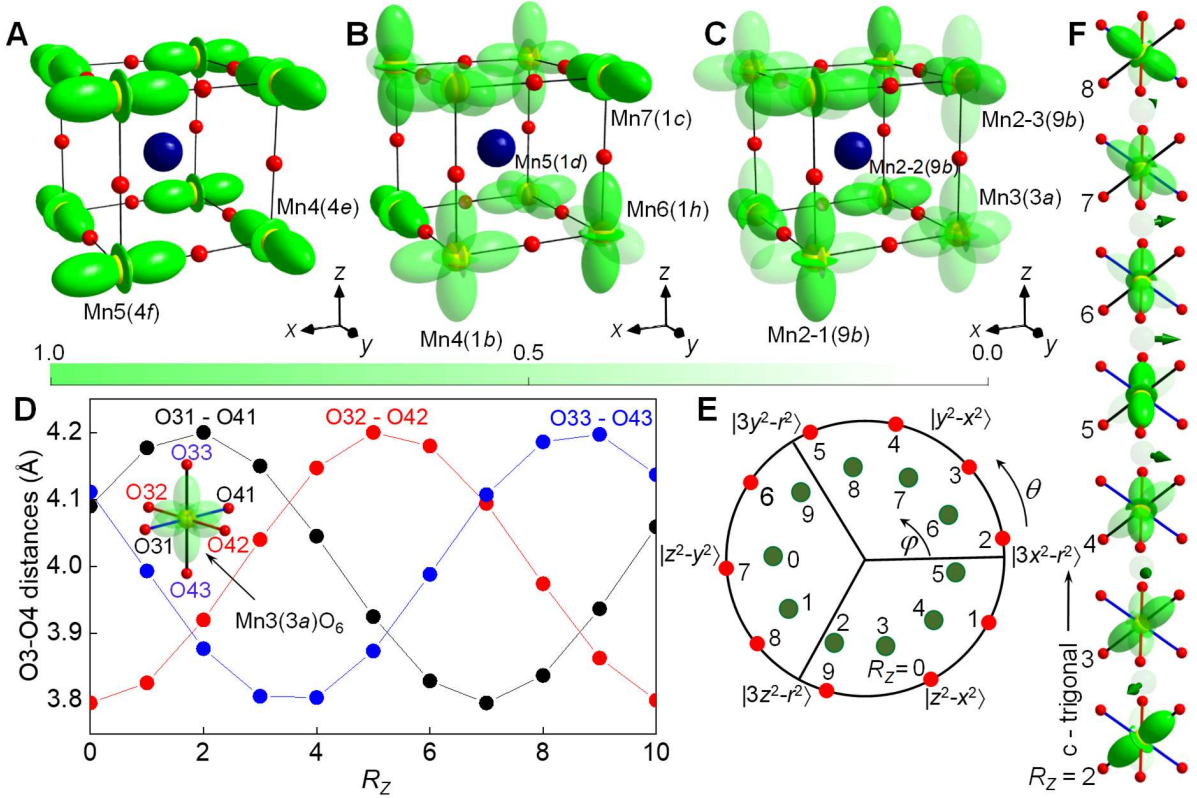

**Fig. 3. Orbital ordering.** Schematic representation of the B-site orbital patterns in the **(A)** monoclinic, **(B)** triclinic and **(C)** trigonal phases of $BiCu_{0.1}Mn_{6.9}O_{12}$. For clarity, the patterns are shown relative to the familiar pseudocubic perovskite unit cell, whose edges are co-aligned with the Cartesian coordinate system $(x,y,z)$ used to define the orbital states. In the triclinic (B) and trigonal (C) phases, the orbital ordering is incommensurate and the orbital state of B-site Mn gradually changes upon propagation through the crystal. The probability of electronic occupancy for different orbitals, averaged over the modulation period is represented by an "orbital density" (degree of transparency). In the triclinic structure the orbital polarization changes between: $|3x^2-r^2\rangle$ and $|3z^2-r^2\rangle$ for Mn4, $|3x^2-r^2\rangle$ and $|3y^2-r^2\rangle$ for Mn5, $|3z^2-r^2\rangle$ and $|z^2-y^2\rangle$ for Mn6 – a pattern that, on each site, is strongly reminiscent of $CaMn_7O_{12}$ *(30)*. The Mn7 site adopts unmodulated fully polarized $|3y^2-r^2\rangle$ state (See fig. S4 for more details). In the trigonal structure, the orbital occupancy of the Mn3 site changes between all three polarized states $|3x^2-r^2\rangle$, $|3y^2-r^2\rangle$ and $|3z^2-r^2\rangle$, making the orbital density wave chiral. The orbital polarization of the Mn2 sites in this structure changes between $|3z^2-r^2\rangle$ and $|z^2-x^2\rangle$ for Mn2-1, $|3x^2-r^2\rangle$ and $|y^2-x^2\rangle$ for Mn2-2, $|3y^2-r^2\rangle$ and $|z^2-y^2\rangle$ for Mn2-3. In addition, the phase of the orbital modulation differs by $+/-2/3\pi$ on these three symmetry-related sites (See fig. S5 for more details). (**D**) Oxygen-oxygen bond distances, used to characterize the anisotropy of the Mn3O$_6$ -octahedra, as a function of the lattice translation $R_Z$ along c-axis in the trigonal phase. (**E**) Orbital states of the Mn3 site, defined by the mixing angle $\theta$ ($\tan\theta = \sqrt{3}[d_{O32-O42} - d_{O33-O43}]/[2d_{O31-O41} - d_{O32-O42} - d_{O33-O43}]$) (red circles), and the displacement of Bi in the *ab*-plane, defined by angle $\varphi$ (angle between the Bi-displacement and projection of the *x*-axis on the ab-plane) (green circles) for different values of the lattice translation $R_Z$. In order to reduce the experimental uncertainty, we exploit the relation $d_{O-O} = 2 \langle d_{B-O}\rangle$ for B-O bonds on either side of a given B site. (**F**) Chiral orbital density wave localized on the Mn3 site (transparency represents the probability of electronic occupation, with lower transparency corresponding to higher probability), and modulated displacements of Bi in the trigonal phase.

# Supplementary Materials for

## Emergent helical texture of electric dipoles

Dmitry D. Khalyavin, Roger D. Johnson, Fabio Orlandi, Paolo G. Radaelli, Pascal Manuel, Alexei A. Belik

Correspondence to: dmitry.khalyavin@stfc.ac.uk

**This PDF file includes:**

    Materials and Methods
    Figs. S1 to S10
    Tables S1 to S5

**Materials and Methods**

Sample preparation

$BiCu_xMn_{7-x}O_{12}$ samples with x = 0.05 and x = 0.1 were prepared from stoichiometric mixtures of $Mn_2O_3$, 'MnO$_{1.839}$', CuO (99.99 %) and $Bi_2O_3$ (99.9999 %), where 'MnO$_{1.839}$' is a commercial 'MnO$_2$' (Alfa Aesar, 99.997 %), whose oxygen content was determined to be MnO$_{1.839}$ (a mixture of $Mn_2O_3$ and $MnO_2$). Single-phase $Mn_2O_3$ was prepared from the commercial 'MnO$_2$' (99.997 %) by heating in air at 923 K for 24 h. The corresponding mixtures were placed in Au capsules and treated at 6 GPa in a belt-type high-pressure apparatus at 1373 K for 1 h (heating time to the desired temperature was 10 min). After the heat treatments, the samples were quenched to room temperature (RT), and the pressure was slowly released. The obtained ceramics had the form of cylindrical pellets with a diameter of 5.5 mm and 3 mm thick, and were of high quality with a typical density of 98.9 % of the theoretical value. The microstructure of the samples was studied using scanning electron microscopy (SEM, JSM-6500F operating at 10 kV) and is demonstrated in Fig. S1. The SEM images confirmed that the samples are practically poreless, dense ceramics with grain size varying between 10 and 30 $\mu$m.

X-ray powder diffraction

X-ray powder diffraction (XRPD) data were collected at RT on a RIGAKU MiniFlex600 diffractometer using CuK$\alpha$ radiation (2$\theta$ range of 10 - 140°, a step width of 0.02°, and a counting speed of 0.5 deg/min). Both samples $BiCu_xMn_{7-x}O_{12}$ with x = 0.05 and 0.1 were found to adopt the identical modulated trigonal phase with very close structural parameters. Synchrotron XRPD data were measured at RT on a large Debye−Scherrer camera at the undulator beamline BL15XU of SPring-8 *(48)*. The intensity data were collected between 4° and 60° at 0.003° intervals in 2$\theta$; the incident beam was monochromatic with $\lambda$ = 0.65298 Å. The samples were packed into Lindemann glass capillaries (inner diameter: 0.1 mm), which were rotated during the measurements. XRPD patterns were analysed by the Rietveld method using JANA2006 program *(23)*.

Neutron diffraction

Neutron powder diffraction measurements were performed at the ISIS pulsed neutron and muon facility of the Rutherford Appleton Laboratory (UK), on the WISH diffractometer located at the second target station *(49)*. The sample of $BiCu_{0.1}Mn_{6.9}O_{12}$ (~ 2 g) was loaded into cylindrical 6 mm diameter vanadium can and measured on warming at the temperature range of 1.5 K < $T$ < 600 K using Oxford instrument cryostat for the measurements below room temperature and hot-stage CCR for the measurements above room temperature. The normalization of the time of flight diffraction data was done using the reduction routines implemented into the Mantid software *(50)*. Rietveld refinement of the crystal structure was performed using the JANA2006 program *(23)*. The structure solution and refinement were assisted by group theoretical calculations done using the ISOTROPY *(21)* and ISODISTORT *(22)* software.

Dielectric measurements

Polarization-field (P-E loop) measurements were performed for two ceramic samples of $BiCu_{0.05}Mn_{6.95}O_{12}$ and $BiCu_{0.1}Mn_{6.9}O_{12}$ by the Positive-Up-Negative-Down (PUND) method. Measurements were performed at $T$ = 77 K such that the electronic conductivity was sufficiently small. Dense pellets of ~ 5 mm diameter were thinned down to a thickness of ~ 0.4 mm, painted with a silver paste electrodes, and mounted within a custom measurement apparatus. Sinusoidal peaks with a 15Hz waveform were employed, separated by 5 s delays. The maximum voltage applied before risk of dielectric breakdown was 1900 V, giving a maximum electric field of 4.75 kV/mm. The charge change upon electric polarization switching was measured directly using an integrating capacitor in a Sawyer-Tower circuit. The P-E loop was found to be reproducible and robust for a range of maximum electric fields, peak widths (frequency) and delay times.

Thermal properties

Differential scanning calorimetry (DSC) curves of powder samples were recorded on a Mettler Toledo DSC1 STAR$^e$ system at a heating/cooling rate of 10 K/min between 130 K and 620 K in sealed Al capsules. Three runs were performed to check the reproducibility, and very good reproducibility was observed.

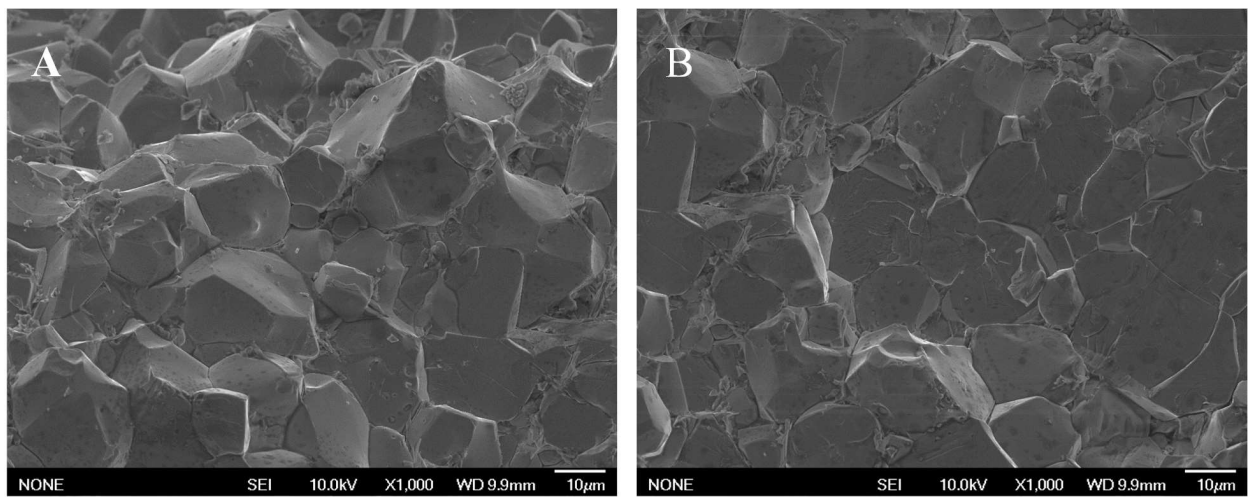

**Fig. S1**
SEM images of fractured surfaces of the $BiCu_xMn_{7-x}O_{12}$ samples with x = 0.05 (A) and x = 0.1 (B).

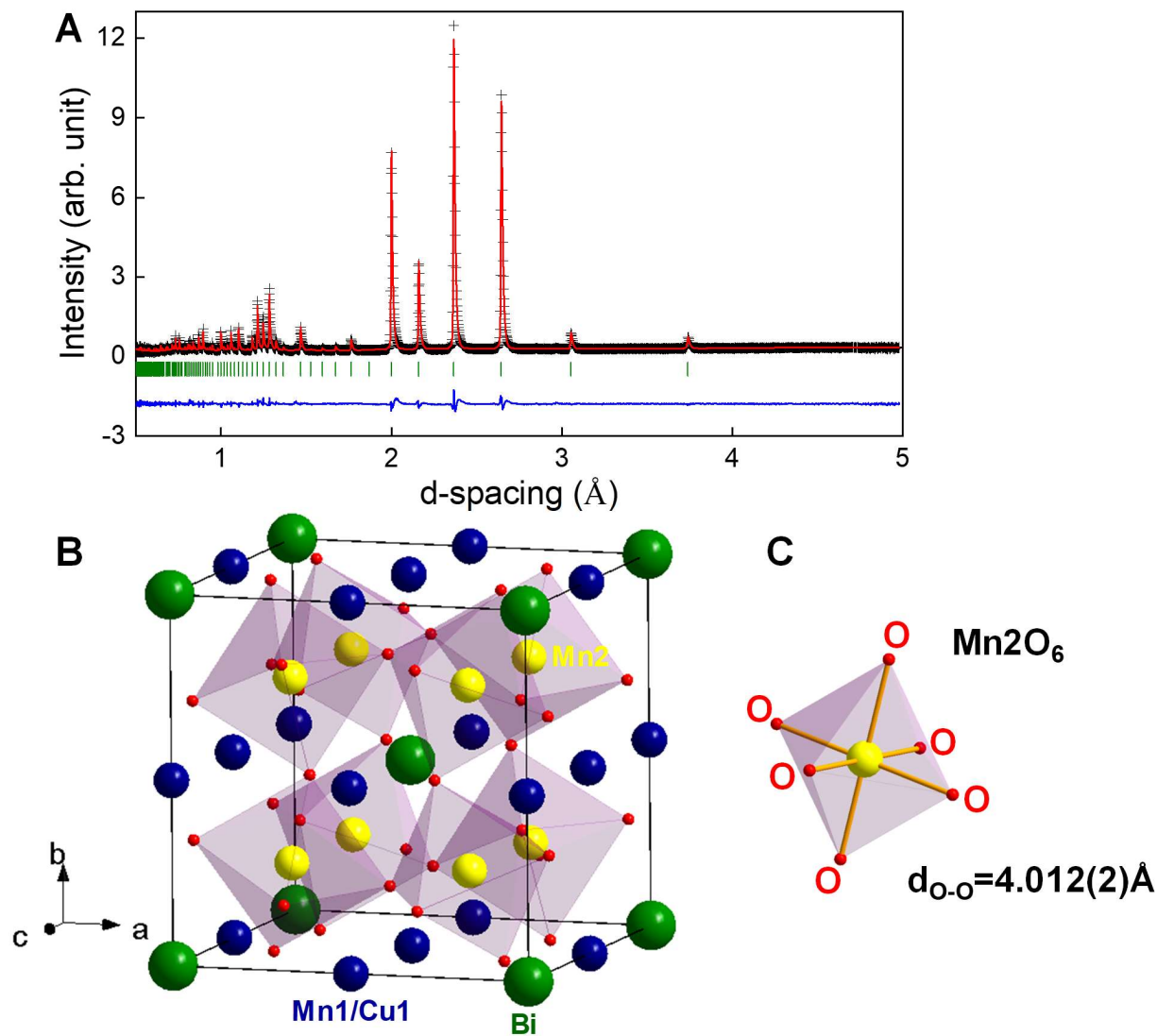

**Fig. S2.**
(**A**) Rietveld refinement of neutron diffraction data (backscattering detectors bank) of $BiCu_{0.1}Mn_{6.9}O_{12}$, collected at $T = 600$ K and refined in the cubic $Im$-3 space group ($R_p = 5.33\%$, $R_{wp} = 5.13\%$). Cross black symbols and solid red line represent experimental and calculated intensities, respectively and the difference between them is given as solid blue line at the bottom of the panel. The green tick marks indicate positions of the Bragg reflections. (**B**) Schematic representation of the cubic unit cell (the structure derives from the prototype cubic perovskite structure $ABO_3$ with the $Pm$-3$m$ space group by A-site cation ordering between Bi and Mn1/Cu1 and in-phase octahedral tilting, the B-site perovskite position is exclusively occupied by Mn (Mn2)). (**C**) Fully isotropic octahedron coordinating B-site Mn.

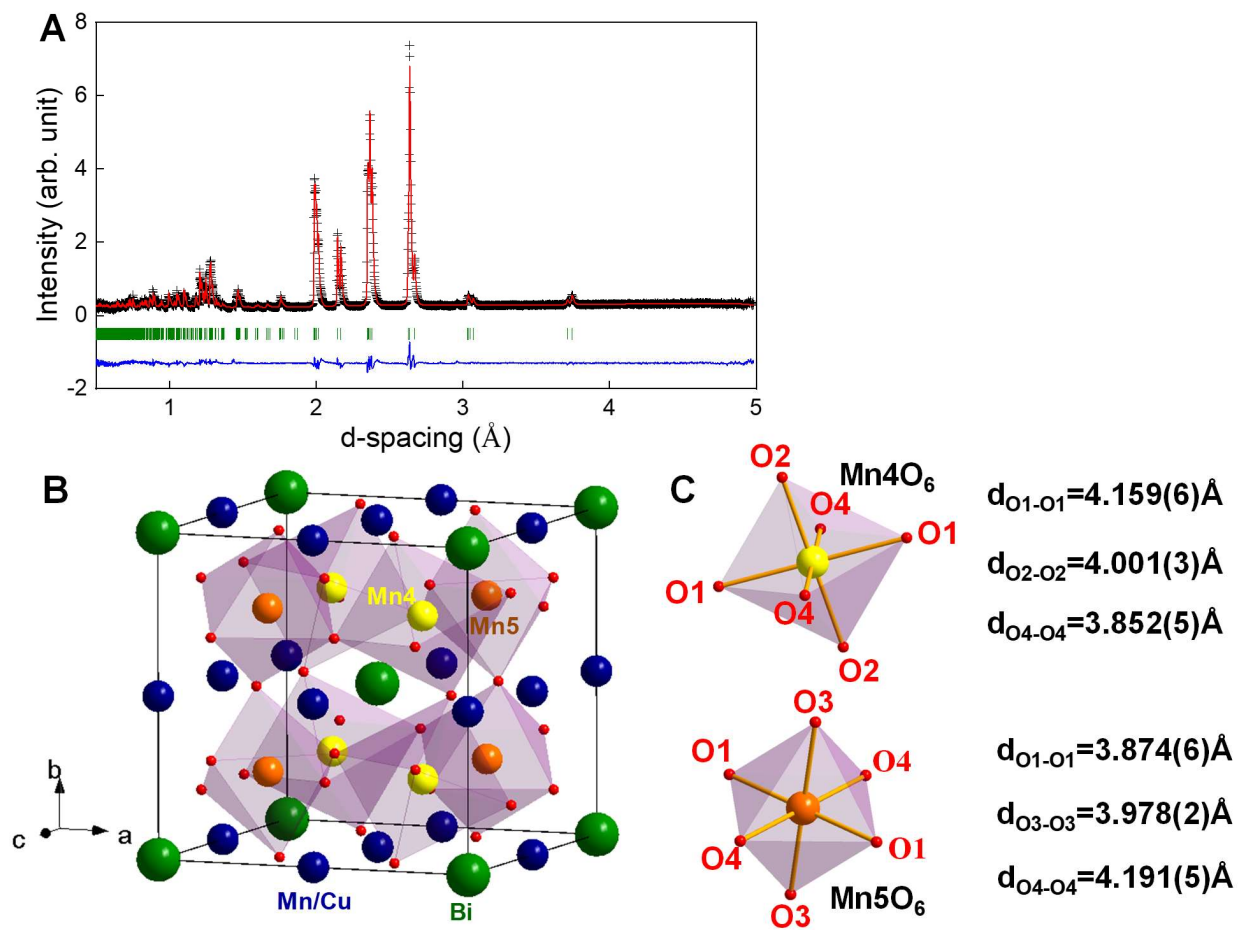

**Fig. S3.**
(**A**) Rietveld refinement of neutron diffraction data (backscattering detectors bank) of $BiCu_{0.1}Mn_{6.9}O_{12}$, collected at $T = 485$ K and refined in the monoclinic $I2/m$ space group ($R_p = 5.57\%$, $R_{wp} = 5.74\%$). Cross black symbols and solid red line represent experimental and calculated intensities, respectively and the difference between them is given as solid blue line at the bottom of the panel. The green tick marks indicate positions of the Bragg reflections. (**B**) Schematic representation of the monoclinic unit cell. (**C**) Octahedra coordinating B-site Mn and list of oxygen-oxygen distances.

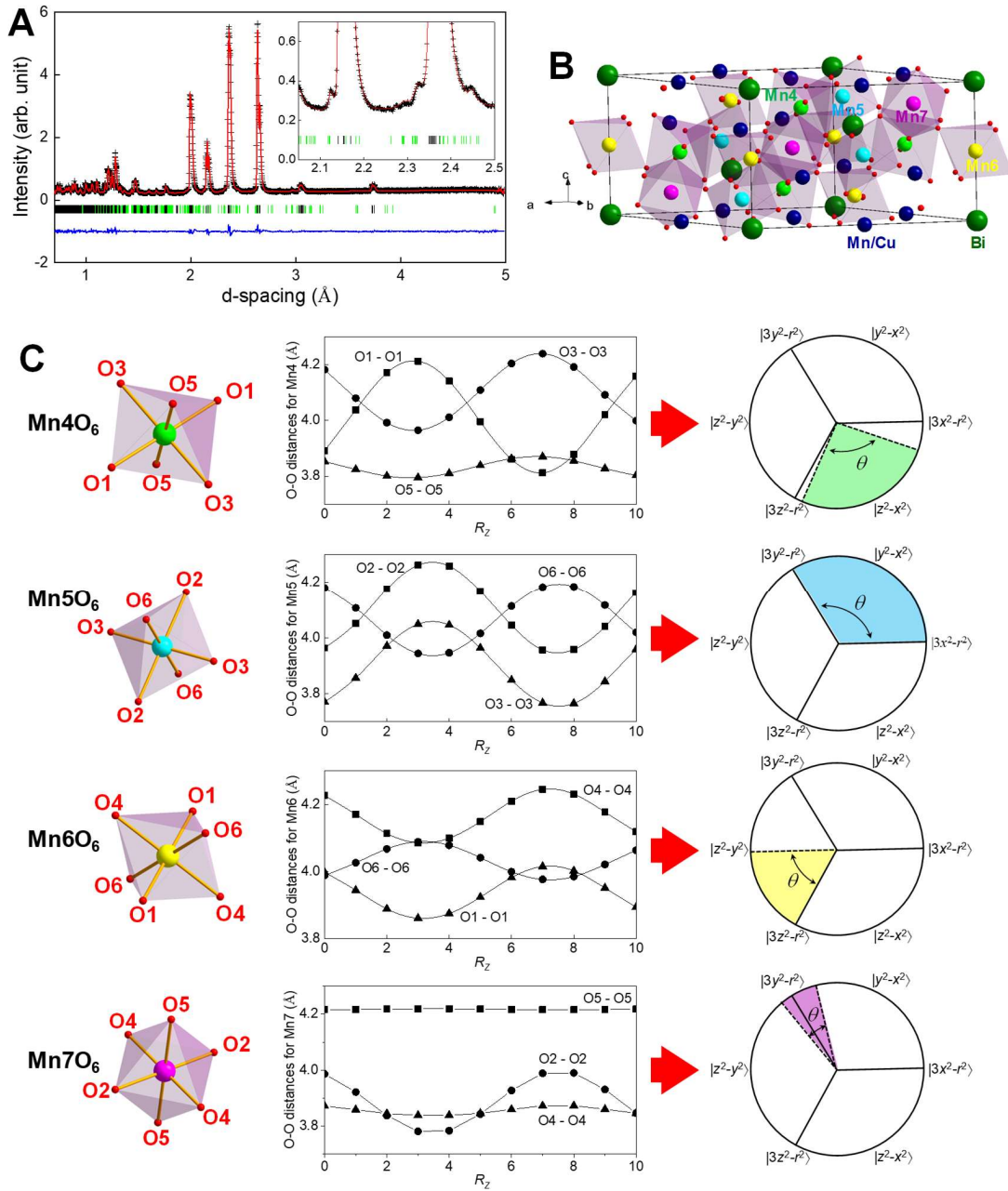

**Fig. S4.**
(**A**) Rietveld refinement of neutron diffraction data (backscattering detectors bank) of $BiCu_{0.1}Mn_{6.9}O_{12}$, collected at $T = 427$ K and refined in the triclinic $R\text{-}1(\alpha\beta\gamma)0$ superspace group ($R_p = 2.91\%$, $R_{wp} = 3.47\%$). Cross black symbols and solid red line represent experimental and calculated intensities, respectively and the difference between them is given as solid blue line at the bottom of the panel. The black and green tick marks indicate positions of the fundamental and satellite Bragg reflections, respectively. (**B**) Schematic representation of the average triclinic unit cell. (**C**) Octahedra coordinating B-site Mn and their oxygen-oxygen bond distances as a function of the lattice translation $R_Z$ along the c-axis. The shaded regions of the corresponding orbital planes show the angular sectors taken by θ through the period of the orbital modulation.

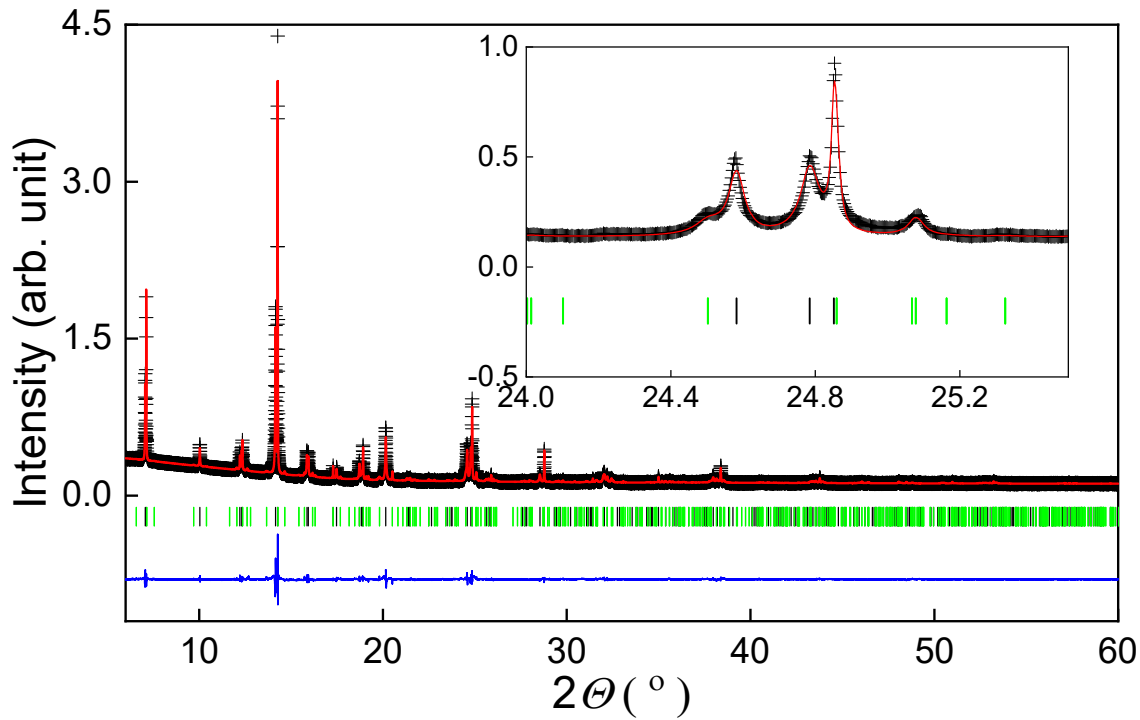

**Fig. S5.**
Rietveld refinement of synchrotron X-ray diffraction data of $BiCu_{0.1}Mn_{6.9}O_{12}$, collected at $T = 300$ K and refined in the trigonal $R3(00\gamma)t$ superspace group ($R_p = 2.86\%$, $R_{wp} = 3.07\%$), using the structural parameters obtained from the neutron diffraction data listed in Table S4. Cross black symbols and solid red line represent experimental and calculated intensities, respectively and the difference between them is given as solid blue line at the bottom of the panel. The black and green tick marks indicate positions of the fundamental and satellite Bragg reflections, respectively.

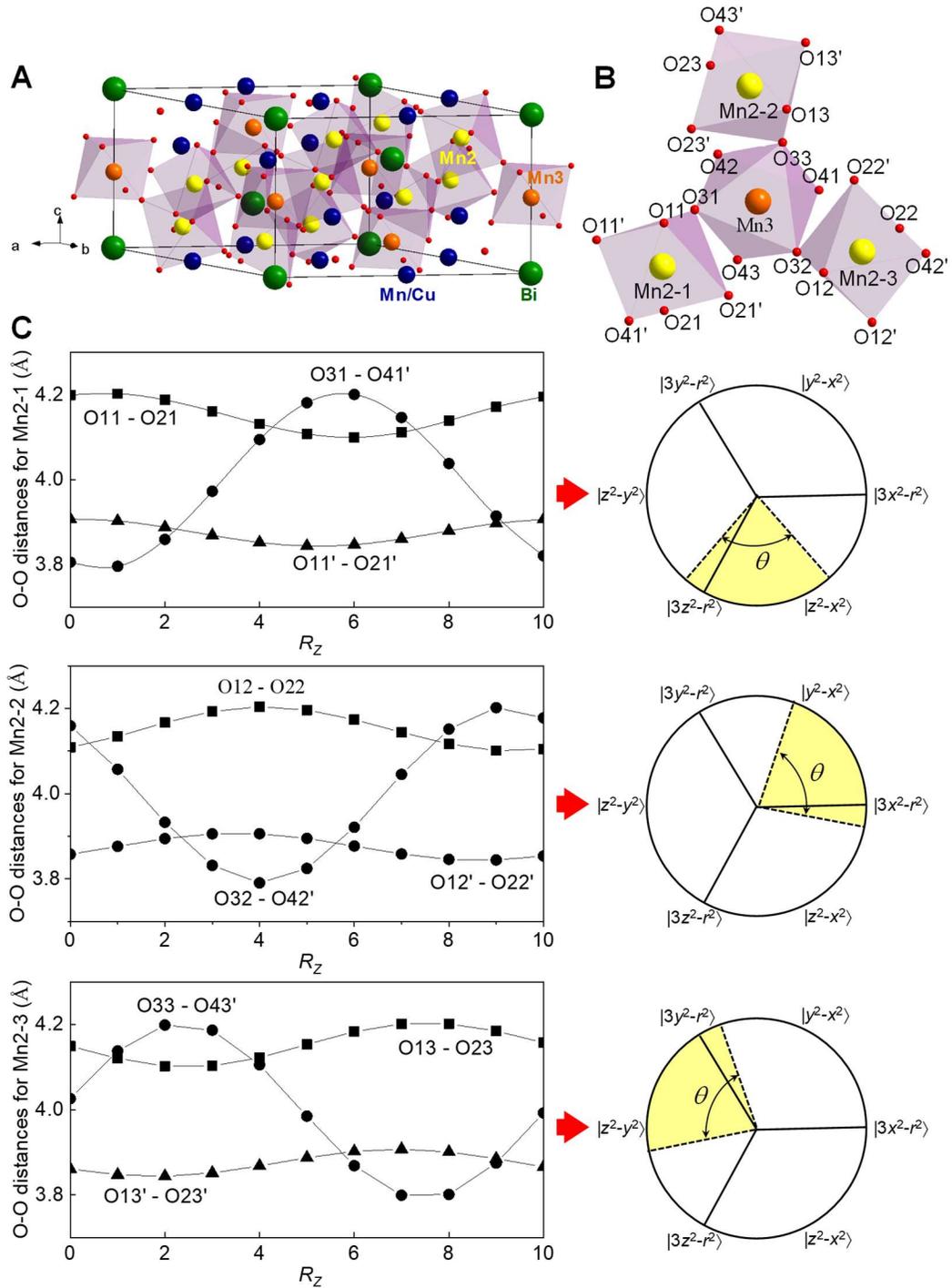

**Fig. S6.**
(**A**) Schematic representation of the average trigonal unit cell. (**B**) Structural fragment showing three symmetry related Mn2-1, Mn2-2 and Mn2-3 sites and oxygen atoms coordinating them. (**C**) Oxygen-oxygen bond distances for the octahedra coordinating Mn2-1, Mn2-2 and Mn2-3, as a function of the lattice translation $R_Z$ along the c-axis. The shaded regions of the corresponding orbital planes show the angular sectors taken by θ through the period of the orbital modulation.

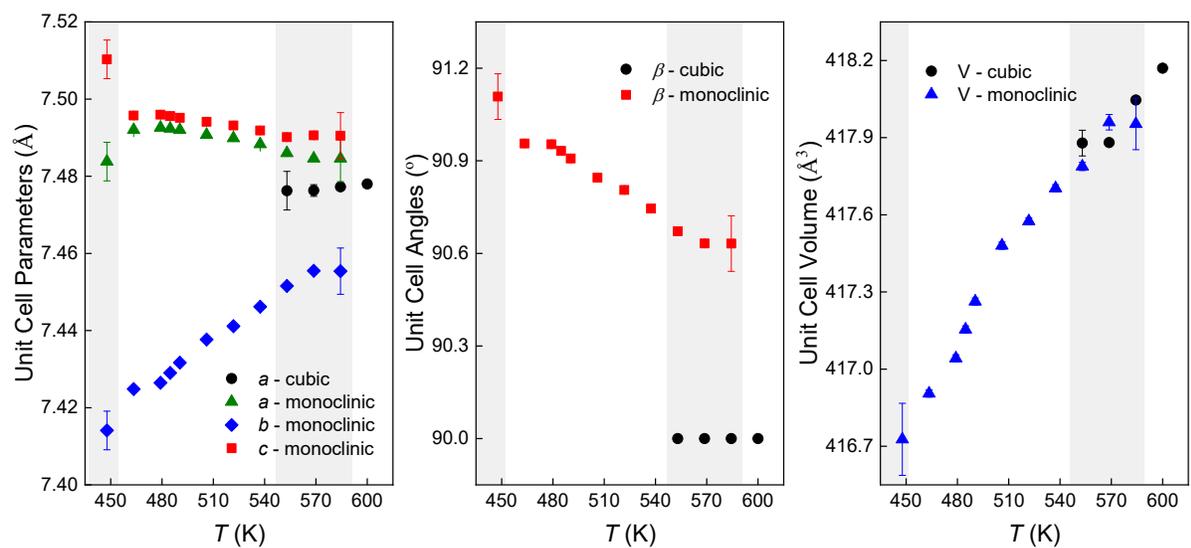

**Fig. S7.**

Unit cell parameters of the high-temperature unmodulated cubic (*Im*-3) and monoclinic (*I*2/*m*) phases as function of temperature for $BiCu_{0.1}Mn_{6.9}O_{12}$, obtained from refinements of the neutron diffraction data (grey shadow indicates regions of the phase coexistence).

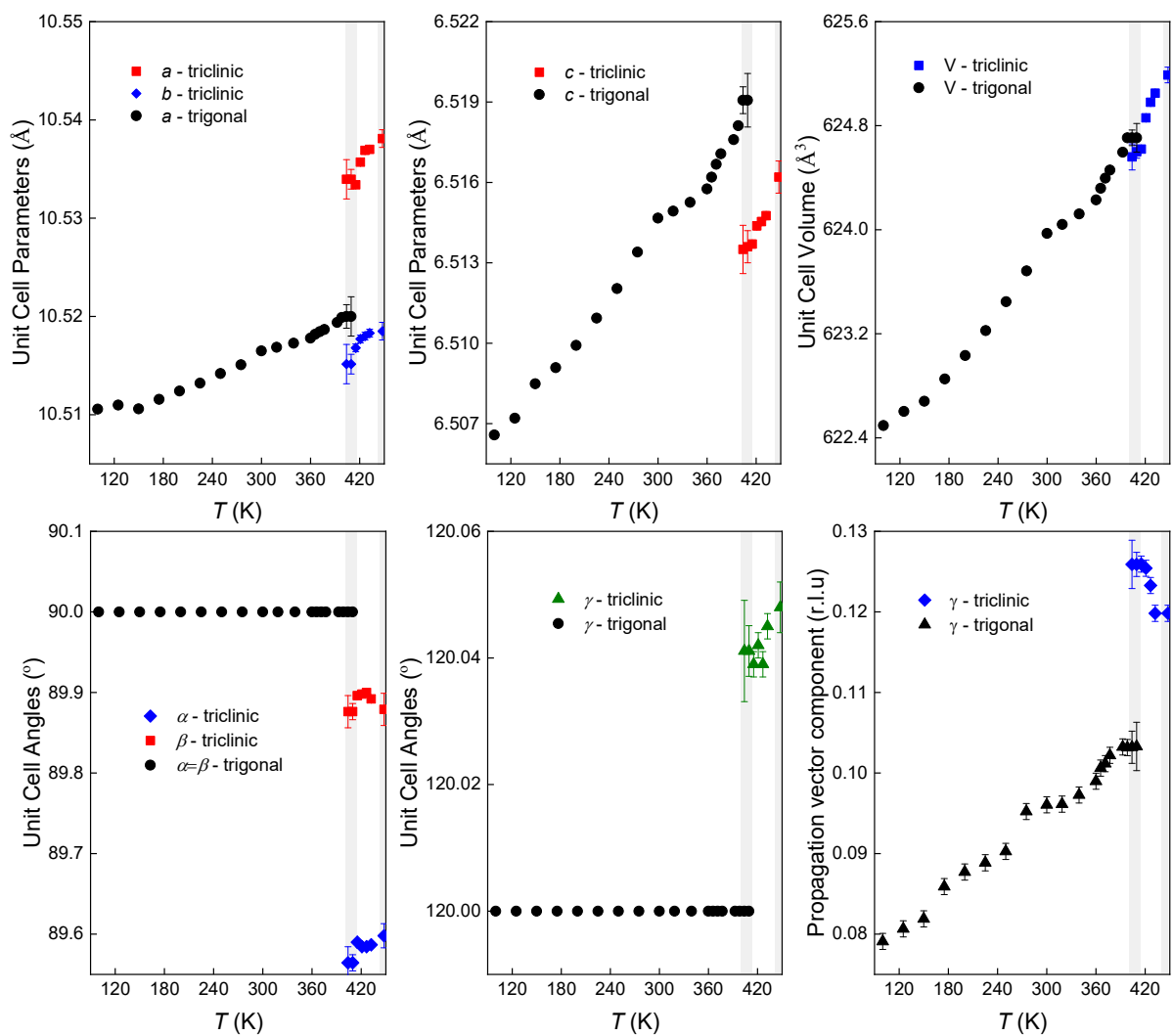

**Fig. S8.**

Unit cell parameters of the low-temperature modulated triclinic ($R$-$1(\alpha\beta\gamma)0$) and trigonal ($R3(00\gamma)t$) phases as function of temperature for BiCu$_{0.1}$Mn$_{6.9}$O$_{12}$, obtained from refinements of the neutron diffraction data (grey shadow indicates regions of the phase coexistence).

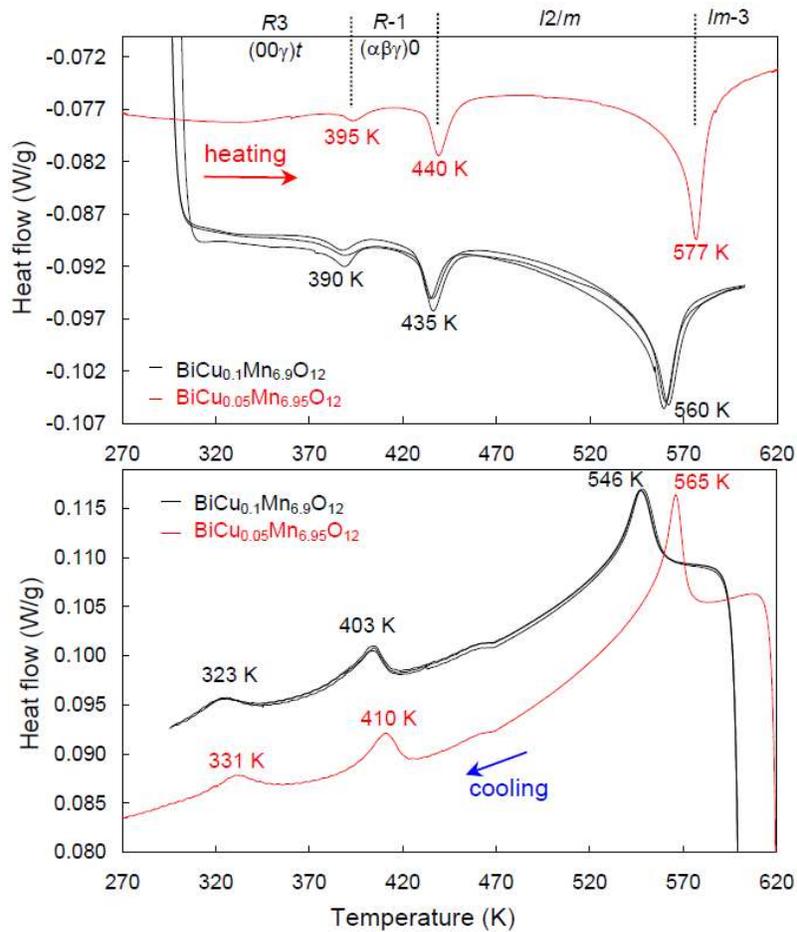

**Fig. S9.**
Differential scanning calorimetry curves on heating (10 K/min) and cooling (10 K/min) for $BiCu_{0.05}Mn_{6.95}O_{12}$ (one run; red curve) and $BiCu_{0.1}Mn_{6.9}O_{12}$ (three runs; black curves). Small kinks near 470 K on cooling curves are instrumental artifacts observed in every measurement.

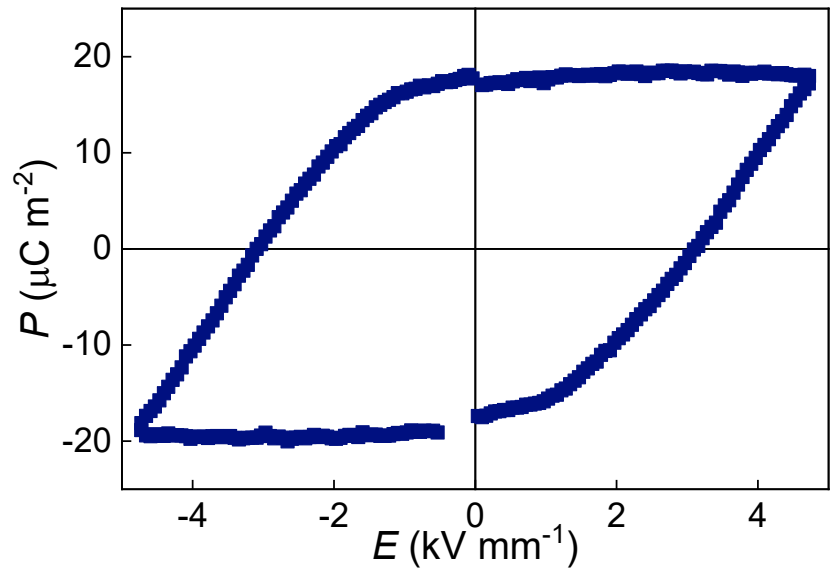

**Fig. S10.**
*P-E* loop measured for BiCu$_{0.1}$Mn$_{6.9}$O$_{12}$ ceramic sample at *T* = 77 K.

**Table S1.**

Atomic coordinates and thermal parameters for $BiCu_{0.1}Mn_{6.9}O_{12}$ at $T = 600$ K (neutron diffraction data), refined in the cubic $Im$-3 space group. Unit cell parameter: $a = 7.47794(4)$ Å and $V = 418.164(4)$ Å$^3$. The ratio between Mn and Cu in the Mn1/Cu1 site was fixed to the nominal chemical composition (0.967/0.033).

| Atom | Site | x | y | z | $U_{iso}$ |
|---|---|---|---|---|---|
| Bi | 2a | 0 | 0 | 0 | 0.056(2) |
| Mn1/Cu1 | 6b | 0 | 0.5 | 0.5 | 0.025(1) |
| Mn2 | 8c | 0.25 | 0.25 | 0.25 | 0.017(1) |
| O | 24g | 0 | 0.3126(2) | 0.1755(2) | 0.0226(7) |

**Table S2.**

Atomic coordinates and thermal parameters for $BiCu_{0.1}Mn_{6.9}O_{12}$ at $T = 485$ K (neutron diffraction data), refined in the monoclinic $I2/m$ space group with the lattice vectors related to the high-temperature cubic structure as: (1,0,0),(0,1,0),(0,0,1) and origin at (0,0,0). Unit cell parameters: $a = 7.4923(1)$ Å, $b = 7.4290(1)$ Å, $c = 7.4956(1)$ Å, $\beta = 90.932(1)°$ and $V = 417.15(1)$ Å$^3$. The ratio between Mn and Cu in the Mn1/Cu1, Mn2/Cu2 and Mn3/Cu3 sites was fixed to the nominal chemical composition (0.967/0.033).

| Atom    | Site | x         | y         | z         | $U_{iso}$  |
|---------|------|-----------|-----------|-----------|------------|
| Bi      | 2a   | 0         | 0         | 0         | 0.054(2)   |
| Mn1/Cu1 | 2c   | 0.5       | 0         | 0         | 0.019(1)   |
| Mn2/Cu2 | 2d   | 0.5       | 0.5       | 0         | 0.019(1)   |
| Mn3/Cu3 | 2b   | 0         | 0.5       | 0         | 0.019(1)   |
| Mn4     | 4e   | 0.75      | 0.25      | 0.75      | 0.0115(8)  |
| Mn5     | 4f   | 0.25      | 0.25      | 0.75      | 0.0115(8)  |
| O1      | 8j   | 0.0092(6) | 0.1756(6) | 0.6876(5) | 0.016(1)   |
| O2      | 4i   | 0.6905(6) | 0         | 0.8284(7) | 0.014(2)   |
| O3      | 4i   | 0.3158(6) | 0         | 0.8197(6) | 0.013(1)   |
| O4      | 8j   | 0.1747(5) | 0.6881(6) | 0.0109(5) | 0.017(2)   |

**Table S3.**
Atomic coordinates of the average structure, modulation and thermal parameters for BiCu$_{0.1}$Mn$_{6.9}$O$_{12}$ at $T$ = 427 K (neutron diffraction data), refined in the triclinic $R$-1($\alpha\beta\gamma$)0 superspace group with the lattice vectors related to the high-temperature cubic structure as: (0,-1,1,0),(-1,1,0,0),(-1/2,-1/2,-1/2,0),(0,0,0,1) and origin at (0,0,0,0). Unit cell parameters: $a$ = 10.5369(2) Å, $b$ = 10.5180(3) Å, $c$ = 6.5145(1) Å, $\alpha$ = 89.585(2)°, $\beta$ = 89.899(2)°, $\gamma$ = 120.039(2)°, V = 624.99(3) Å$^3$ and the components of the modulation vector: $\alpha$ = -0.0037(8), $\beta$ = 0.026(1) and $\gamma$ = 0.1233(8). The ratio between Mn and Cu in the Mn1/Cu1, Mn2/Cu2 and Mn3/Cu3 sites was fixed to the nominal chemical composition (0.967/0.033). $A^1_i$ and $B^1_i$ ($i=x,y,z$) are the Fourier coefficients of the first harmonic ($n$=1) of the displacive modulation function: $u_{i,j,l}(r_{j,l} \cdot k_{HT}) = \sum_{n=0}^{\infty} A^n_{i,j} \cos(2\pi n[r_{j,l} \cdot k_{HT}]) + B^n_{i,j} \sin(2\pi n[r_{j,l} \cdot k_{HT}])$, where $r_{j,l}$ indicates the position of the $j$-th atom of the average structure in the $l$-th unit cell.

| Atom | Site | x | y | z | $U_{iso}$ |
|---|---|---|---|---|---|
| $A^1_i$ | | $A^1_x$ | $A^1_y$ | $A^1_z$ | |
| $B^1_i$ | | $B^1_x$ | $B^1_y$ | $B^1_z$ | |
| Bi | 1a | 0 | 0 | 0 | 0.046(2) |
| | | 0 | 0 | 0 | |
| | | 0.032(2) | 0.001(3) | 0.008(3) | |
| Mn1/Cu1 | 1e | 1/3 | 1/6 | -1/3 | 0.023(1) |
| | | 0 | 0 | 0 | |
| | | 0.012(2) | -0.004(4) | 0.003(6) | |
| Mn2/Cu2 | 1g | 1/6 | 1/3 | -2/3 | 0.023(1) |
| | | 0 | 0 | 0 | |
| | | 0.012(2) | -0.004(4) | 0.003(6) | |
| Mn3/Cu3 | 1f | -1/6 | 1/6 | -1/3 | 0.023(1) |
| | | 0 | 0 | 0 | |
| | | 0.012(2) | -0.004(4) | 0.003(6) | |
| Mn4 | 1b | 2/3 | 5/6 | -1/6 | 0.0130(8) |
| | | 0 | 0 | 0 | |
| | | 0.016(1) | 0 | 0 | |
| Mn5 | 1d | -5/6 | -2/3 | -1/6 | 0.0130(8) |
| | | 0 | 0 | 0 | |
| | | 0.016(1) | 0 | 0 | |
| Mn6 | 1h | 1/3 | 2/3 | 1/6 | 0.0130(8) |
| | | 0 | 0 | 0 | |
| | | 0.016(1) | 0 | 0 | |
| Mn7 | 1c | -1/2 | -1/2 | -1/2 | 0.0130(8) |
| | | 0 | 0 | 0 | |
| | | 0.021(1) | 0 | 0 | |
| O1 | 2i | 0.172(1) | 0.520(1) | 0.339(1) | 0.0126(4) |
| | | -0.005(2) | 0.002(2) | 0.009(3) | |
| | | 0.025(2) | 0.008(2) | 0.017(2) | |
| O2 | 2i | -0.293(1) | -0.4011(9) | -0.572(1) | 0.0126(4) |
| | | 0.007(2) | 0.008(2) | -0.003(2) | |
| | | 0.023(2) | -0.005(2) | 0.000(3) | |
| O3 | 2i | 0.741(1) | 0.7814(8) | 0.094(1) | 0.0126(4) |
| | | 0.005(2) | -0.005(2) | 0.000(3) | |
| | | 0.0254(2) | 0.011(2) | 0.005(3) | |
| O4 | 2i | 0.489(1) | 0.650(1) | 0.342(1) | 0.0126(4) |
| | | 0.005(2) | -0.001(2) | -0.011(3) | |
| | | 0.021(2) | 0.005(2) | 0.012(2) | |
| O5 | 2i | -0.1124(9) | 0.296(1) | -0.567(1) | 0.0126(4) |
| | | 0.000(2) | -0.002(2) | 0.000(2) | |
| | | 0.006(2) | 0.009(2) | 0.001(3) | |
| O6 | 2i | -0.342(1) | 0.171(1) | -0.340(1) | 0.0126(4) |
| | | 0.007(2) | 0.005(2) | 0002(3) | |
| | | 0.026(2) | 0.010(2) | -0.007(3) | |

**Table S4.**

Atomic coordinates of the average structure, modulation and thermal parameters for $BiCu_{0.1}Mn_{6.9}O_{12}$ at $T = 300$ K (neutron diffraction data), refined in the trigonal $R3(00\gamma)t$ superspace group with the lattice vectors related to the high-temperature cubic structure as: (0,-1,1,0),(-1,1,0,0),(-1/2,-1/2,-1/2,0),(0,0,0,1) and origin at (0,0,0,0). Unit cell parameters: $a = 10.51635(8)$ Å, $c = 6.5148(1)$ Å, $V = 623.97(1)$ Å$^3$ and the component of the modulation vector: $\gamma = 0.0974(5)$. The ratio between Mn and Cu in the Mn1/Cu1 site was fixed to the nominal chemical composition (0.967/0.033). $A^1_i$ and $B^1_i$ ($i=x,y,z$) are the Fourier coefficients of the first harmonic ($n=1$) of the displacive modulation function: $u_{i,j,l}(r_{j,l} \cdot k_{LT}) = \sum_{n=0}^{\infty} A^n_{i,j} \cos(2\pi n [r_{j,l} \cdot k_{LT}]) + B^n_{i,j} \sin(2\pi n [r_{j,l} \cdot k_{LT}])$, where $r_{j,l}$ indicates the position of the $j$-th atom of the average structure in the $l$-th unit cell.

| Atom | Site | x | y | z | $U_{iso}$ |
|---|---|---|---|---|---|
| $A^1_i$ |  | $A^1_x$ | $A^1_y$ | $A^1_z$ |  |
| $B^1_i$ |  | $B^1_x$ | $B^1_y$ | $B^1_z$ |  |
| Bi | 3a | 0 | 0 | 0 | 0.031(3) |
|  |  | 0.0208(6) | 0.042(1) | 0 |  |
|  |  | -0.036(1) | 0 | 0 |  |
| Mn1/Cu1 | 9b | 0.487(2) | 0.004(1) | 0.010(3) | 0.013(2) |
|  |  | 0.015(2) | 0.016(2) | 0.000(4) |  |
|  |  | -0.002(2) | 0.021(2) | -0.020(2) |  |
| Mn2 | 9b | 0.502(4) | 0.001(5) | 0.503(5) | 0.006(2) |
|  |  | 0.018(2) | 0.027(2) | 0.004(3) |  |
|  |  | -0.005(2) | 0.011(3) | 0.005(3) |  |
| Mn3 | 3a | 0 | 0 | 0.483(4) | 0.006(2) |
|  |  | 0.019(9) | 0.009(4) | 0 |  |
|  |  | 0.001(3) | 0.017(3) | 0 |  |
| O1 | 9b | 0.219(1) | 0.259(1) | 0.096(4) | 0.0105(7) |
|  |  | 0.013(2) | 0.021(2) | 0.003(3) |  |
|  |  | -0.020(2) | 0.001(2) | -0.006(2) |  |
| O2 | 9b | -0.224(1) | -0.264(1) | -0.100(4) | 0.0105(3) |
|  |  | 0.010(2) | 0.013(2) | 0.000(3) |  |
|  |  | -0.024(2) | 0.003(2) | 0.003(3) |  |
| O3 | 9b | 0.676(1) | 0.178(1) | 0.008(3) | 0.0105(3) |
|  |  | 0.021(2) | 0.022(2) | -0.002(2) |  |
|  |  | -0.009(2) | 0.011(3) | 0.013(2) |  |
| O4 | 9b | -0.686(1) | -0.188(1) | -0.003(3) | 0.0105(3) |
|  |  | 0.017(2) | 0.018(2) | -0.002(2) |  |

| | -0.018(3) | 0.018(3) | -0.005(2) |

**Table S5.**
Bond distances extracted from the structural parameters listed in Tables S1-S4.

| Cubic ($T$ = 600 K) $Im$-3 | | Monoclinic ($T$ = 485 K) $I2/m$ | | Triclinic ($T$ = 427 K) Average $R$-1 | | Trigonal ($T$ = 300 K) Average $R3$ | |
|---|---|---|---|---|---|---|---|
| Bond | Distance (Å) | Bond | Dist. (Å) | Bond | Dist. (Å) | Bond | Dist. (Å) |
| Bi-O | 2.681(1)×12 | Bi-O1 | 2.682(4)×4 | Bi-O1 | 2.70(2)×2 | Bi-O1 | 2.63(2)×3 |
| | | Bi-O2 | 2.633(5)×2 | Bi-O2 | 2.68(3)×2 | Bi-O2 | 2.66(2)×3 |
| | | Bi-O3 | 2.744(5)×2 | Bi-O3 | 2.61(2)×2 | Bi-O3 | 2.66(2)×3 |
| | | Bi-O4 | 2.662(4)×4 | Bi-O4 | 2.74(3)×2 | Bi-O4 | 2.75(2)×3 |
| | | | | Bi-O5 | 2.63(3)×2 | | |
| | | | | Bi-O6 | 2.69(3)×2 | | |
| Mn1/Cu1-O | 1.920(1)×4 | Mn1/Cu1-O2 | 1.937(5)×2 | Mn1/Cu1-O3 | 1.95(3)×2 | Mn1/Cu1-O1 | 2.08(4)×1 |
| | | Mn1/Cu1-O3 | 1.916(4)×2 | Mn1/Cu1-O4 | 1.90(2)×2 | Mn1/Cu1-O2 | 1.78(4)×1 |
| | | Mn2/Cu2-O1 | 1.918(4)×4 | Mn2/Cu2-O1 | 1.94(4)×2 | Mn1/Cu1-O3 | 1.92(2)×1 |
| | | Mn3/Cu3-O4 | 1.915(4)×4 | Mn2/Cu2-O2 | 1.94(3)×2 | Mn1/Cu1-O4 | 1.92(2)×1 |
| | | | | Mn3/Cu3-O5 | 1.92(3)×2 | | |
| | | | | Mn3/Cu3-O6 | 1.88(3)×2 | | |
| Mn2-O | 2.0061(4)×6 | Mn4-O1 | 2.080(4)×2 | Mn4-O1 | 2.01(3)×2 | Mn2-O1 | 2.05(4)×1 |
| | | Mn4-O2 | 2.000(2)×2 | Mn4-O3 | 2.04(2)×2 | Mn2-O1 | 1.95(3)×1 |
| | | Mn4-O4 | 1.926(4)×2 | Mn4-O5 | 1.92(2)×2 | Mn2-O2 | 2.11(4)×1 |
| | | Mn5-O1 | 1.937(4)×2 | Mn5-O2 | 2.05(2)×2 | Mn2-O2 | 1.95(3)×1 |
| | | Mn5-O3 | 1.989(2)×2 | Mn5-O3 | 1.95(3)×2 | Mn2-O3 | 2.02(4)×1 |
| | | Mn5-O4 | 2.095(4)×2 | Mn5-O6 | 2.03(2)×2 | Mn2-O4 | 1.98(4)×1 |
| | | | | Mn6-O1 | 1.97(2)×2 | Mn3-O3 | 2.09(4)×3 |
| | | | | Mn6-O4 | 2.07(3)×2 | Mn3-O4 | 1.91(4)×3 |
| | | | | Mn6-O6 | 2.02(3)×2 | | |
| | | | | Mn7-O2 | 1.94(2)×2 | | |
| | | | | Mn7-O4 | 1.93(3)×2 | | |
| | | | | Mn7-O5 | 2.11(2)×2 | | |